# Enhanced Photocurrent in solution processed electronically coupled CdSe Nanocrystals thin films

Hareesh Dondapati, Duc Ha and A. K. Pradhan*

Center for Materials Research, Norfolk State University, 700 Park Avenue,

Norfolk, VA 23504

**Abstract**

We have demonstrated the fabrication of highly continuous and smooth CdSe semiconductor films containing self-assembled nanocrystals (NCs) using a simple, low cost solution-processed deposition technique. The impact of thermal annealing and ethanedithiol (EDT) treatment on oleate capped CdSe NCs films is illustrated. Post deposition EDT treatment enhances strong electron coupling between NCs by reducing the inter-particle distance, which enhances four orders of magnitude of photocurrent in the *pn*-device. Mild thermal annealing of NC films cause large redshift and significant broadening. Our findings suggest that NCs with short-range organic ligands are suitable for high-performance Thin-Film-Transistors (TFTs) and next generation high-efficiency photovoltaics.

*Corresponding author's email: apradhan@nsu.edu

Solution processed semiconductor nanocrystal (NC) thin films have become one of the most promising building blocks for optoelectronic devices.[1] Extensive research has already been reported on epitaxial growth of NC thin films using ultra high vacuum based techniques, however, together with wet chemical approach, self-assembled, and solution processed NCs electronic and optoelectronic devices offer low cost, large device area and physical flexibility.[2] In particular, the size and shape dependent fine quantum space confinement regime drives the development of NC-based optoelectronic technologies. Besides their unique properties, such as bandgap tunability, multiple exciton generation[3,4] and self-assembly, inexpensive device fabrication can be advantageous through drop casting, and spin coating. The potential advantage of colloidal NC solids is the deposition on any surface and their electronic properties can be controlled by adjusting NC composition, size, shape, monodispersity, capping ligand, and surface chemistry.[5,6] For most device applications of semiconductor NCs, the electronic coupling between NCs must be sufficient in order to facilitate efficient motion of charges between each NCs. Typically, conventional synthesis of high-quality semiconductor NCs utilizes long hydrocarbon tails that form insulating barriers around each NC and inhibit exciton separation and decrease charge-carrier mobility in close-packed NC films.[7-9]

In order to convert NC solids from insulators to conductors, insulating organic ligand shells that must be either shortened or removed[10] needs to be introduced. A common way to enhance the charge carrier mobility is to introduce smaller organic ligands, so that the electronic coupling between adjacent NCs becomes higher.[11-14] Recent reports from Nozik, et al[10-14], Luther et al[13,14] and Talpin complexes (MCCs)[16], Octadecylamine (ODA) and $SnS_4$ capping layers[17,18] triggered the need of detailed study on optical and electrical properties of NC thin films treated both thermally and chemically. Each NC synthesis procedure leads to a specific long rage organic ligand at the end. Detailed study on specific organic ligand is very limited. The aim of the present work is to study the detailed optical and electrical characterization of oleate capped CdSe NC thin films on various substrates via thermal annealing and thiol-based chemical treatments. Figure 1(a) shows schematic illustrating of ligand-exchange process, showing that EDT removes the electrically insulating native ligand molecules. Figure 1 (b) demonstrates the fabrication process of self-assembled nanocrystal thin film devices.

Two key process parameters whose optimization allowed us to enhance few orders of magnitude of photocurrent in CdSe thin films. Reducing the distance between the NCs and keeping it within the electron tunneling limit is very crucial to enhance the photocurrent. Here we report how various thermal treatments and 1, 2-ethanedithiols (EDT) chemical treatment significantly enhanced the photocurrent without decreasing quantum size. A significant enhancement in photocurrent was demostrated on post-deposition EDT treatment of NC CdSe films on p-Si as well.

CdSe NCs were synthesised using a typical wet chemistry method.[16,7] Standard airfree techniques were used throughout. The typical synthesis procedure is described briefly as follows: first cadmium myristate solution was prepared in a three-neck flask by heating a mixture of 0.154 g of CdO with 0.58 g of myristic acid in 10 ml of 1-octadecene at 240 °C followed by the addition of 64 ml of 1-octadecene and then degassing the solution at 90 °C under vacuum for 1h. Then 0.048 g of Se powder was introduced to the reaction mixture after the solution was cooled to room temperature. Subsequently, the mixture was heated to 240 °C for three min and a solution of 0.2 mL of oleic acid and 2 mL of oleylamine in 8 mL of 1-octadecene was added dropwise to the reaction mixture. The solution was maintained at 240 °C for 1 h, then cooled to room temperature. The NCs were purified by adding ethanol or acetone to the crude solution, transferred to a glovebox and redispersed in anhydrous hexane. All films are fabricated in a nitrogen filled glove box. For optical characterization studies at different annealing temperatures, originally oleate capped 200 nm thick NC films in hexane (~12 mg ml$^{-1}$) were deposited onto glass substrates using spin coat technique, and then annealed in N$_2$ at 100–400 $^o$C for 10 min. Au/NCs/*p*-Si and Au/NCs/ITO structures, as shown in figure 1(b) were fabricated for electrical characterization. In order to understand charge injection and transport properties a 200 nm thick NC films are deposited on ITO/glass substrates. To study the photovoltaic effect simple *pn* junction devices are fabricated on highly Boron doped passivated *p*-Si (100) wafers with the resistivity of 1 to 10 Ω-cm. On ITO and *p*- Si, films are fabricated using a layer-by-layer (LbL) dip coating method. Top Au metal electrode was deposited using electron beam evoporation technique. In this LbL method, a layer of NCs is deposited onto the surface by dip coating from a hexane solution and then washed in 0.01 M 1,2-ethanedithiol (EDT) in acetonitrile to remove the electrically insulating oleate ligands.

The synthesis process yielded highly monodispersed NCs with a consistently narrow size distribution, as seen in the transmission electron microscopy (TEM). Our TEM results showed in Fig. 2 (a) revealed that these NCs seem to be well dispersed, suggesting that they are well passivated by oleate organic ligands. NCs are approximately spherical shape with diameters ranging from 3 to 4 nm. It is also clear that one can observe from high resolution TEM image showed in Fig. 2(b) that the lattice fringes are visible along different crystal axes, confirming their single-crystalline nature. The inset in Fig. 2 (a) indicates high crystallinity of the NCs with a lattice parameter of 0.25 nm.

To understand the thermal stability, the normalized absorption spectra of NC thin films as a function of annealing temperature is shown in Fig. 3. It is noted that both absorption and emission spectra were taken at room temperature. The bottom left inset shows the absorption and emission spectrum of NCs with original organic ligands in solution exhibits well-resolved features with a first absorption peak at 561 nm, which corresponds to NCs with diameter of 3.2 nm close to the size determined from the TEM image. This deep in absorbance spectrum is due to strong exitonic effect in CdSe NCs. The photoluminescence spectra of NCs show a peak at 515 nm corresponding to an optical band gap of 2.15 eV. The thermal treatment causes absorption feature that exhibits a large redshift and significant broadening. The first excitonic peak shifts by 16 nm, from 561 to 577 nm. The first excitonic peak is very well resolved, indicating that the quantum confinement is still present upon annealing up to 300 $^o$C. The identical features of the NCs disappear after annealing at 350 and 400 $^o$C indicating that NCs are no longer distinctly visible at higher temperatures, because thermal treatments at higher treatments can also lead to sintering of NC films. The broadening of absorption peak and red shift can be attributed to a decreased separation between annealed NCs.[19]

On the other hand, upper right inset shows that the treatment with EDT in acetonitrile causes the first excitonic peak of NC thin films to red shifts by 8 nm. This red shift can be explained with the help of: (a) modification of size of the NCs due to thiols on the surface of the NCs, (b) enhanced electronic coupling via decrease in the NCs spacing, could lead to a Mott-type insulator- to- conductor transition[20], and (c) change of dielectric constant of the medium surrounding the NCs after oleate loss.

In effort to decrease the inter-NC spacing and also to remove the surface oleate compound, the NC thin films on silicon substrates were treated with EDT. Another batch of samples was annealed in the presence of $N_2$ at 100 – 350 $^o$C for 10 min. Fig. 4 compares the Fourier transform infrared (FTIR) spectra of as made CdSe NC thin film with EDT treated sample. One can observe that the organic bonds are present in the as deposited thin films. The peaks labeled at 2920 cm$^{-1}$ and 2852 cm$^{-1}$ in the CdSe - oleate complex represent the anti-symmetric and symmetric stretching of the C–H bonds along the alkyl chain of the oleic acid chain. The peak observed at 2956 cm$^{-1}$ is associated with the $CH_3$ stretching and the peaks at 1462 cm$^{-1}$ and 722 cm$^{-1}$ arise from the deformation and rocking vibration of the $CH_2$ bond respectively. In the case of EDT treated sample the absence of S-H stretch signal at around 2550 cm$^{-1}$ indicates that the oleate ligands have been completely removed. This can be further confirmed from the loss of the peak at 2956 cm$^{-1}$. The inset in Fig. 4 compares the FTIR spectra of annealed samples. We can clearly see that the intensity of C-H stretch decrease as a function of annealing temperature. Significant reduction in intensities is usually attributed to a decreased separation between annealed NCs.[16] Furthermore, at these temperatures NCs start to grow based on the previous report on PbSe thin film solids.[10] This suggests that the possible formation of inorganic polycrystalline CdSe thin films. This is also further supported by our absorption spectra shown in Fig. 3.

Fig. 5 shows field emission scanning electron microscopy (FESEM) images of NC films on silicon substrates. A smooth and compact surface with high-density of nanocrystals was observed. It confirms that spin coat technique can be efficient to produce very smooth and continuous thin films. We also confirmed that films do possess certain degrees of cracking upon chemical treatment as a result of oleate loss.

We have investigated the effect of thermal annealing on the charge transport and electron coupling behavior of NC thin films that are originally oleate capped, and were subjected to annealing in $N_2$ at different temperatures prior to the deposition of Au. Our results demonstrate that it is a tradeoff between quantum confinement and charge injection capability at CdSe NCs and ITO interface. Our absorption spectra revealed that the cutoff annealing temperature is around 300 °C before the organic ligands decompose and NC films behave like a bulk CdSe semiconductor, where the first absorption peak close to 690 nm. As both ITO and CdSe are *n*-type, we expect to see a liner response with respect to the applied voltage across the electrodes as a function of thermal treatment. These are nearly symmetric, indicating similar electron injection efficiencies from the ITO and Au electrodes (see Fig. 6(a)). Due to the organic barrier surrounded by NCs electron injection can only be possible through quantum tunneling phenomena. Due to highly insulating nature of NCs the conductivity of films annealed up to 200 °C is unchanged and unresponsive to the applied voltage, however NC films which are treated with EDT showed a strong coupling among them. As a consequence, current-voltage (I-V) curve possess an ohmic behavior as shown in Fig. 6 (a). These data indicate that the thermal treatment just alone is not sufficient for significant removal of oleate for strong electron coupling and efficient electron injection.

Fig. 6 (b) shows the current-voltage (I-V) characteristics of *pn* junction device, which was fabricated using a 200 nm thick EDT treated CdSe NC film on *p*-Si substrate, and Au served as electrodes. The measurements are taken in the dark and under illumination with white light conditions using CUDA lamp of intensity 1 Watt/cm$^2$ on 1 cm$^2$. Our dark measurement demonstrates Schottky barrier junction behavior at the *p*-Si/NC interface and drives exciton dissociation and charge transport. Furthermore, remarkable increase in photocurrent, which is about four orders of magnitudes, has been observed under illumination with light due to strong electron coupling caused by EDT treatment. Based on our measurements while not shown here the solar cell parameters are expected to be very high.

In summary, we have demonstrated, a simple, cost effective tool for the fabrication of optoelectronic device comprising of self-assembled solution processed NC thin film solids followed by short-range organic ligand exchange process on various substrates. The high-resolution TEM and optical studies demonstrate the high-quality and mono-disperse nanocrystals. The FTIR results suggest the successful ligand exchange process. We fabricated EDT treated CdSe NCs on *p*-Si. Four order of photocurrent achieved due to strong electron coupling among CdSe NCs through EDT treatment. Our findings also suggest a range of applications, since tunable optical bandgap and electrical properties of CdSe NC solids make them suitable for high performance optoelectronic devices such as next generation thin film transistors (TFTs) and heterojunction solar cells without using Si.

Acknowledgements: This work is supported by the NSF-CREST (CNBMD) Grant No. HRD 1036494, and partially by DoD (CEAND) Grant No. W911NF-11-1-0209 (US Army Research Office). The authors would like thank Olga V. Trofimova, and R. Brandt for their help in characterizing our samples.

Figure Captions

Fig. 1. (a) Schematics of organic ligands exchange process, and (b) demonstrates the fabrication process of self-assembled nanocrystal thin film devices., Au/CdSe NCs/SiO$_2$/p-Si and CdSe NCs/glass device structures.

Fig. 2. (a) Low-resolution TEM image (b) High-resolution TEM (HRTEM) image of 3 – 4 nm oleate capped CdSe NCs. The HRTEM image inset of Fig. 1(a) indicates high crystallinity of the NCs with a lattice parameter of 0.25 nm.

Fig. 3. Room temperature absorption spectra of NC films on glass substrates as a function thermal treatment. Bottom left inset is absorption and emission spectra of colloidal NCs in hexane. Upper right inset is with the effect of EDT treatment.

Fig. 4 Comparison between FTIR spectra of as made and EDT treated NC films on silicon substrate. The inset shows FTIR spectra of NC thin films as a function of thermal treatment.

Fig. 5 (a) and (b) FESEM images of NC thin films on silicon substrates treated with EDT.

Fig. 6 I-V Characteristics of (a) ITO/CdSe NCs/Au structures as a function of annealing temperature and (b) EDT treated CdSe NCs on *p*-Si using a 100 nm SiO$_2$ insulating layer.

Figure 1

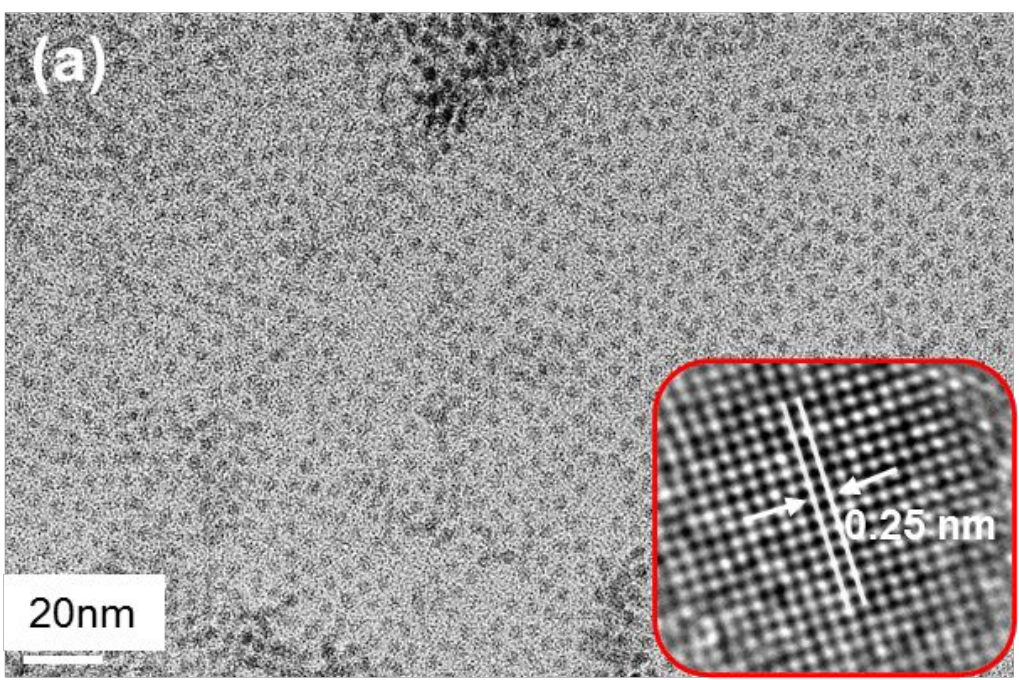
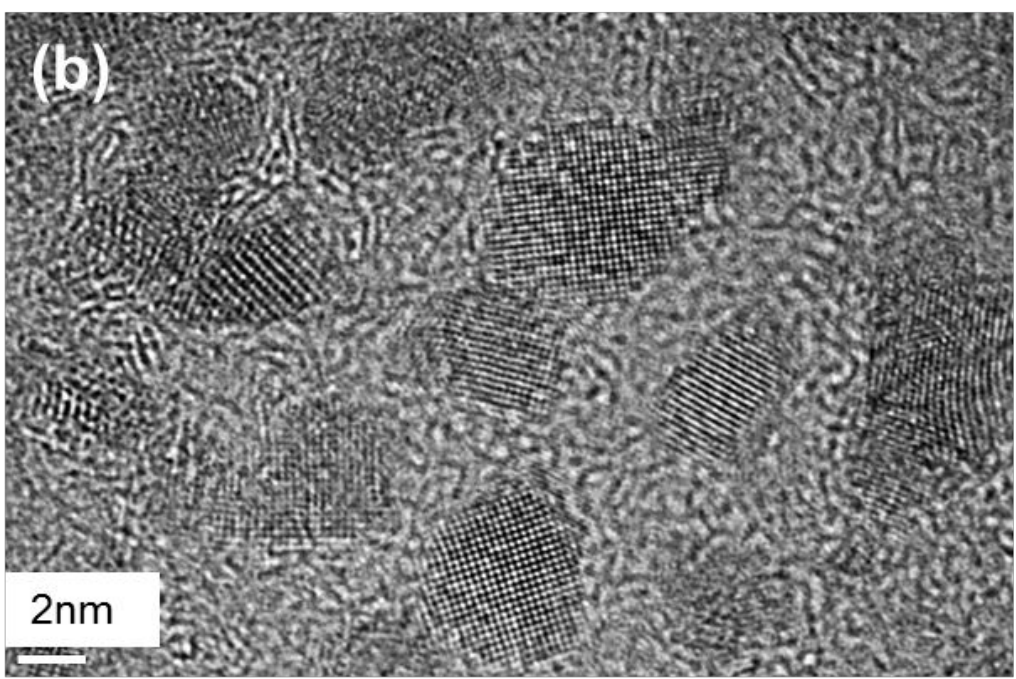

Figure 2

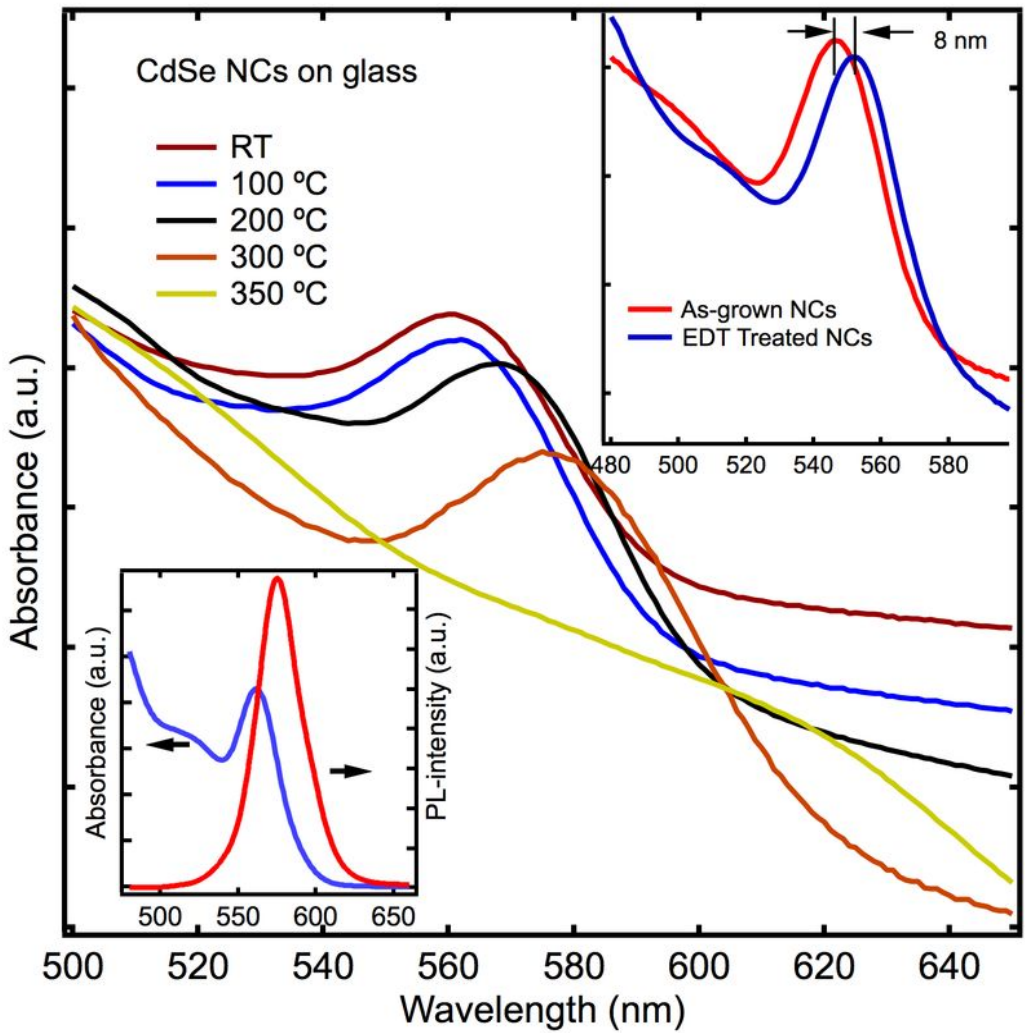

Figure 3

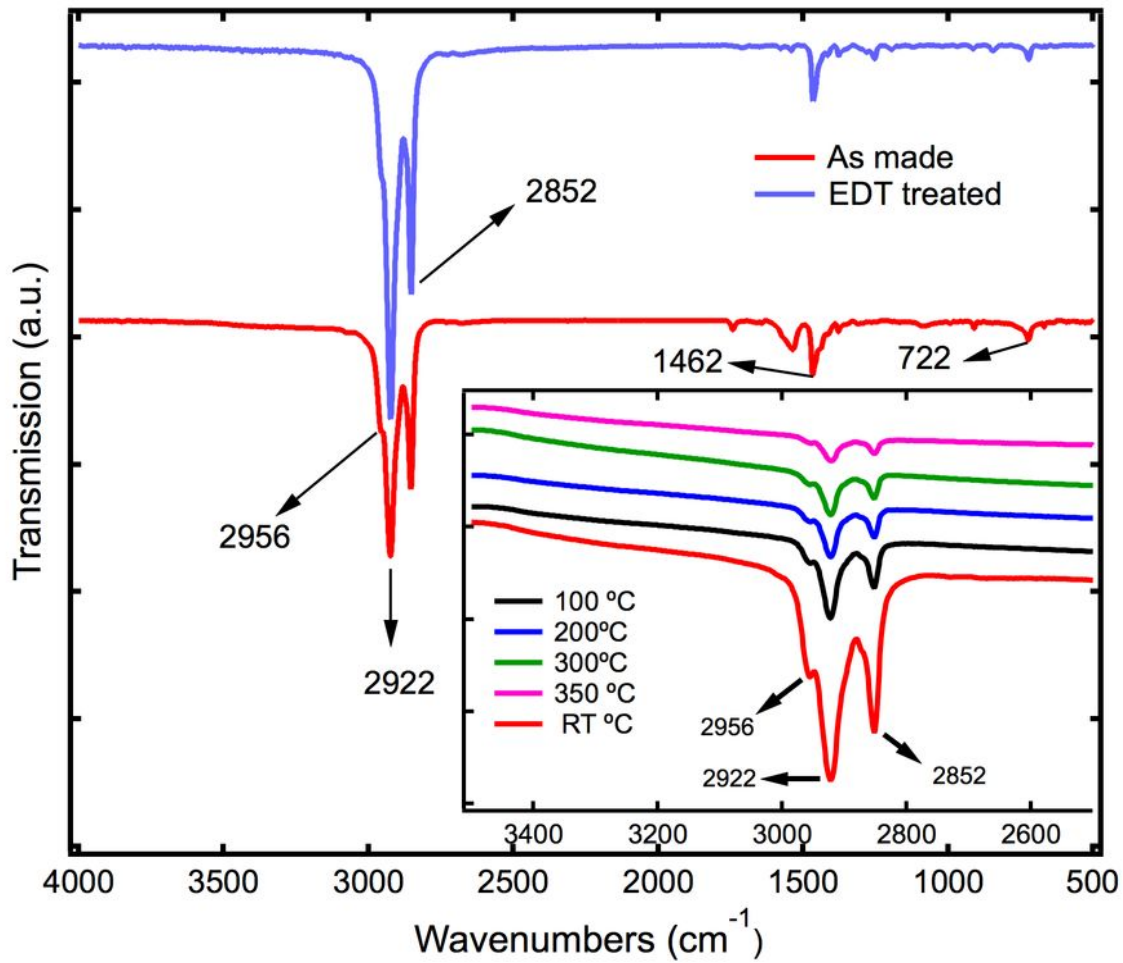

Figure 4

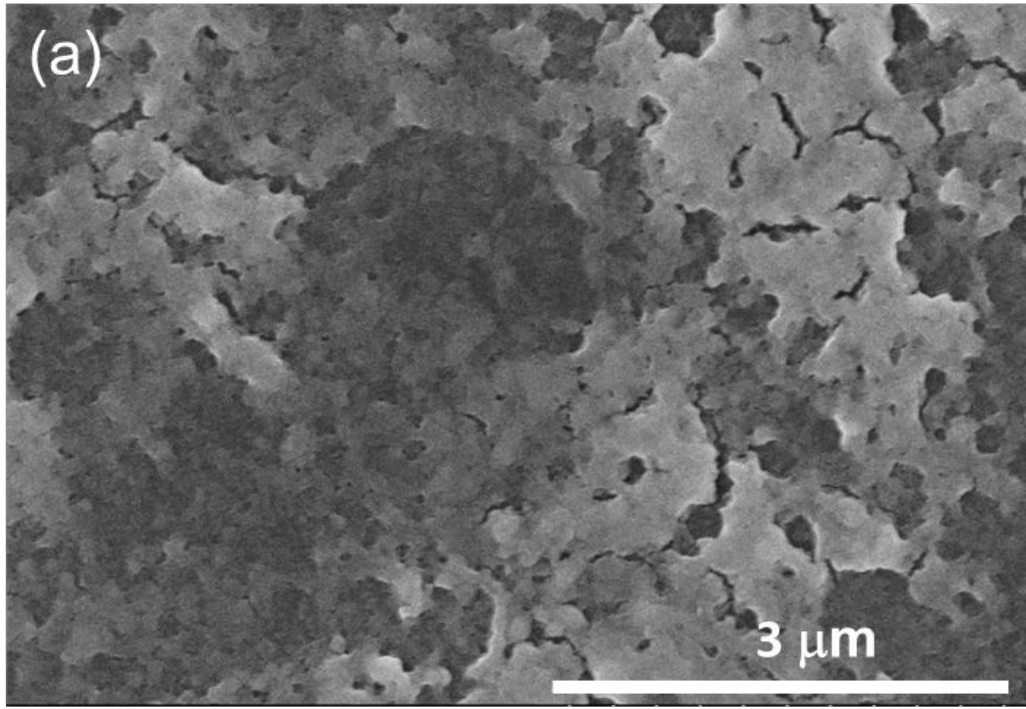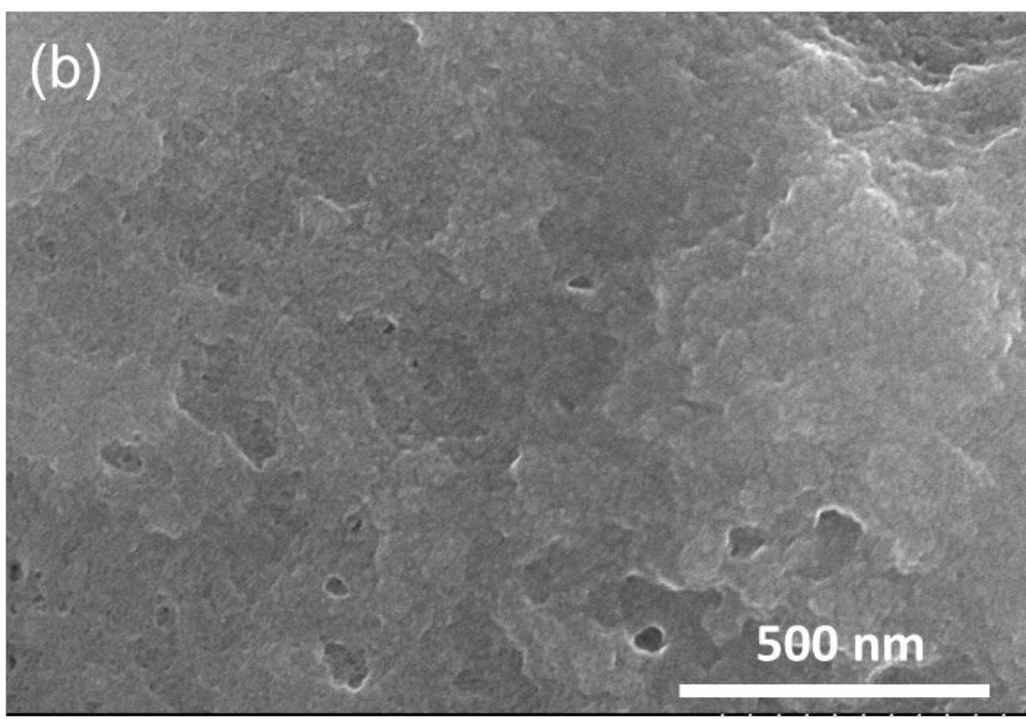

Figure 5

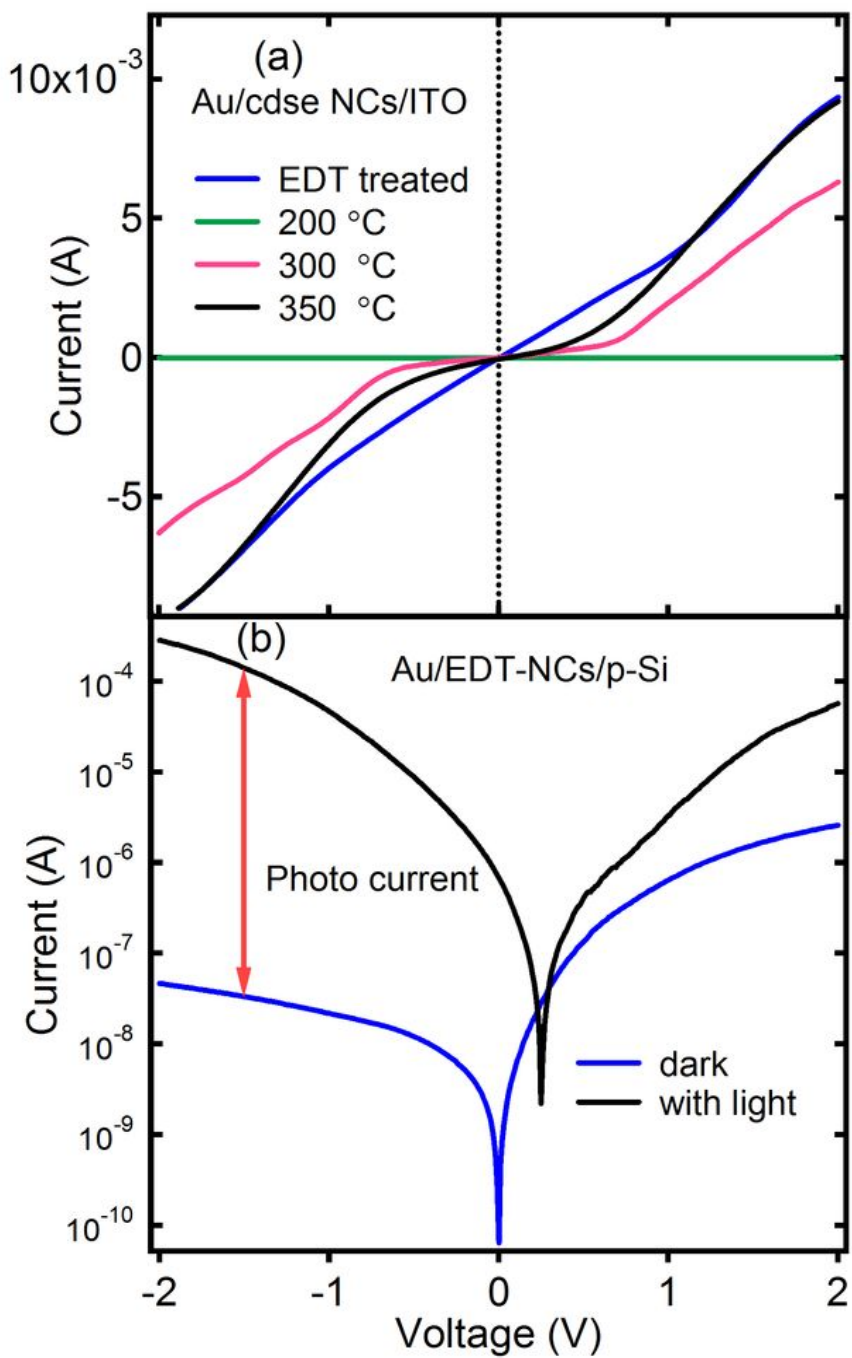

Figure 6